\documentclass[]{aastex62}
\usepackage{amsmath}

\usepackage{xspace}

\graphicspath{{./}{figures/}}

\received{may 1, 2021}
\revised{may 7, 2021}
\accepted{\today}
\submitjournal{ApJ}

\shorttitle{NIR spectroscopy of UC\ion{H}{2} regions in W51A}
\shortauthors{Barbosa et al.}

\begin{document}

\title{Near-Infrared Spectroscopy of Ultracompact \ion{H}{2} regions in W51A with NIFS/ALTAIR\footnote{Based on observations obtained at the Gemini Observatory (processed using the Gemini IRAF package v1.14), which is operated by the Association of Universities for Research in Astronomy, Inc., under a cooperative agreement with the NSF on behalf of the Gemini partnership: the National Science Foundation (United States), National Research Council (Canada), CONICYT (Chile), Ministerio de Ciencia, Tecnolog\'{i}a e Innovaci\'{o}n Productiva (Argentina), Minist\'{e}rio da Ci\^{e}ncia, Tecnologia e Inova\c{c}\~{a}o (Brazil), and Korea Astronomy and Space Science Institute (Republic of Korea).}}

\correspondingauthor{Cassio Barbosa}
\email{cbarbosa@fei.edu.br}

\author[0000-0002-4922-0552]{Cassio L. Barbosa}
\affiliation{Centro Universit\'ario da FEI, Dept. de F\'{\i}sica.\\
Av. Humberto Alencar de Castelo Branco, 3972 \\
S\~ao Bernardo do Campo - SP, CEP 09850-901, Brazil}
\nocollaboration

\author[0000-0002-0284-0578]{Felipe Navarete}
\affiliation{SOAR Telescope/NSF's NOIRLab\\ Avda Juan Cisternas 1500, 1700000, La Serena, Chile}
\affiliation{Universidade de S\~ao Paulo, Instituto de Astronomia, Geof\'{\i}sica e Ci\^{e}ncias Atmosf\'{e}ricas\\
Rua do Mat\~ao, 1226, S\~ao Paulo - SP, CEP 05508-090, Brazil}
\nocollaboration

\author[0000-0002-8622-4237]{Robert D. Blum}
\affiliation{Vera C. Rubin Observatory/NSF's NOIRLab\\
950  N.  Cherry  Ave, Tucson, AZ 85719, USA}
\nocollaboration

\author[0000-0002-7978-2994]{Augusto Damineli}
\affiliation{Universidade de S\~ao Paulo, Instituto de Astronomia, Geof\'{\i}sica e Ci\^{e}ncias Atmosf\'{e}ricas\\
Rua do Mat\~ao, 1226, S\~ao Paulo - SP, CEP 05508-090, Brazil}
\nocollaboration

\author{Peter S. Conti}
\affiliation{In memoriam\\ JILA, University of Colorado, Boulder, CO 80309-0440, USA)}
\nocollaboration



\begin{abstract}
W51A is the most active star formation region of the Giant \ion{H}{2} region W51. It harbors the two massive proto-clusters W51e and W51\,IRS2, which are very rare in the Galaxy. We aim to identify the new born massive stars and UCHII regions  to derive its distance and age. We performed IFU observations with NIFS+ALTAIR of nine targets in the W51A sub-region. The distance modulus was obtained using the spectral classification in the $K$-band and a reddening law appropriate to the inner Galactic plane. We derived the distance and the spectral types for five of them, ranging from O8 to O9.5, similar to those derived from radio continuum data, except for two sources that we assigned somewhat a later spectral type. We included another seven objects with precise spectral classification from other works, which allowed us to better constrain the distance estimate. Our spectrophotometric distance d=4.80\,$\pm$\,1.27\,kpc is in good agreement with those derived from the Galactic rotation model and trigonometric  parallaxes, placing the region near the tangent point of the Sagittarius arm. We conclude that the stars studied in this work have an age spread of 1.5-4 Myr, substantially older than thought to date.
\end{abstract}

\keywords{(ISM:) \ion{H}{2} --- 
stars: massive --- stars: formation --- infrared: ISM --- infrared: stars}

\section{Introduction\label{intro}}

W51 is a massive star-forming complex located near the Carina-Sagittarius arm tangential point which has a kinematic distance of 5.5\,kpc \citep{kolp03}. Due to its location and its vantage point, the distance to W51 is a matter of discussion \citep[e.g.,][]{lys08}. The complex actually comprises the entire W51 Giant Molecular Cloud (GMC) fragmented into four main clusters: G49.58$-$00.38, G49.50$-$0.4 (also known as W51\,North and W51A; hereafter W51A), G49.2$-$0.3 and G48.9$-$0.3. The W51 complex is one of the most intense infrared sources in the sky \citep{kumar04}, and may represent a local case for the starburst phenomenon. For this reason, the complex has been the subject of recent multi-wavelength studies in the near-infrared (NIR) \citep{bik19}, mid-infrared (MIR) \citep{barb16, lim19}, sub-millimeter \citep{ginsb16m} and radio wavelengths \citep{ginsb16}.

The clusters within the W51 GMC show evidence of ongoing massive star formation at different rates. W51A is the youngest and the most active cluster  \citep{oku00, bik19}, with a large number of Ultracompact \ion{H}{2} (UC\ion{H}{2}) regions identified by \citet{meh94} and \citet{ginsb16}. The stellar population of W51A was studied by \citet{oku00}, \citet{kumar04}, and by \citet{lys08} who spectroscopically identified four O stars (within the range O4 - O7.5). More recently, \citet{bik19} reported the identification of another four late O/early B  stars in this cluster based on the analysis of NIR spectroscopy.

\citet{lim19} presented a study of the regions W51A and G49.4$-$0.3 with unprecedented details in the MIR (at 20 and 37\,\micron), confirming the very young nature of both clusters. In their study, 47 point-like sources at the MIR were identified as massive young stellar objects (MYSOs), but only 21 were confirmed as the counterparts of radio sources identified by \citet{meh94} and \citet{ginsb16}. According to the latter authors, the fact that more than 50\% of their sample is still radio quiet indicates that the majority of the MYSOs in W51 have not reached the UC\ion{H}{2} phase yet, placing them at even earlier evolutionary phases, such as Hypercompact \ion{H}{2} (HC\ion{H}{2}) regions or even hot cores \citep{church02}.

The present work concentrates on the investigation of nine UC\ion{H}{2} regions within W51A, and the paper is organized as follows: Section\,\ref{obs} describes the details of the observational procedures and the data reduction.
In Sect.\,\ref{results} we present the spectroscopic results and classification of the NIR sources, including where appropriate, previous information already published and so available to us.
In Sect.\,\ref{distance} we present a new spectrophotometric distance to W51A, and in Sect\,\ref{discuss} we discuss our results, comparing them with previous works. Finally, we summarize our conclusions in Sect.\,\ref{summary}.

\section{Observations and data processing}
\label{obs}

\subsection{Target selection and observations}
\label{target_selection}

The sources observed in this work were selected among the brightest UC\ion{H}{2} regions within W51A, observed with the Very Large Array at 6-cm by \citet{meh94}. 
The 6-cm map was matched with the {3.6\,\micron} image taken with the \textit{Spitzer}/IRAC camera \citep{fazio04} to identify NIR counterparts of the UC\ion{H}{2} regions presented by \citet{barb16}, using a maximum {3\arcsec} offset between the radio and the IR positions.

Finally, the position of the NIR counterparts was refined using the $JHK_{S}$ UKIDSS  Galactic Plane Survey images \citep{luc08}. 
Table\,\ref{logs} presents the details of UC\ion{H}{2} regions and the NIR counterparts identified in the procedure described above.

\setlength{\tabcolsep}{3pt}
\begin{deluxetable*}{ccrrcrccc}[ht]
\tablecaption{Observation log of NIR counterpart candidates of the UC\ion{H}{2}. \label{logs}}
\tablecolumns{11}
\tablewidth{0pt}
\tablehead{
\colhead{UC\ion{H}{2}} &
\colhead{Spectral} &
\colhead{RA} &
\colhead{Decl.} &
\colhead{UKIDSS} &
\colhead{$K_{S}$} &
\colhead{Obs. date} &
\colhead{Exp. Time} &
\colhead{Spat. Res.} \\
\colhead{Region} & 
\colhead{Type\tablenotemark{a}} & 
\colhead{(J2000)} & 
\colhead{(J2000)} & 
\colhead{designation} & 
\colhead{(mag)} & 
\colhead{(UTC)} & 
\colhead{(seconds)} & 
\colhead{(\arcsec)}
}
\startdata
W51a	&	O6	&	19:23:29.67	&	+14:31:35.0	&	J192329.67+143135.0	&	10.81 &	2015-06-30	&	400	&	0.17	\\
W51b$_{1}$	&	O8.5	&	19:23:34.72	&	+14:32:05.2	&	J192334.72+143205.2	&	11.77	&	2015-06-30	&	800	&	0.21	\\
W51b$_{2}$	&	B0	&	19:23:35.88	&	+14:31:28.8	&	J192335.88+143128.8	&	12.06 &	2015-07-01	&	800	&	0.25	\\
W51c$_{1}$	&	O5	&	19:23:41.03	&	+14:29:27.0	&	J192341.03+142927.0	&	11.64 &	2015-06-30	&	800	&	0.18	\\
W51e$_{7}$	&	B0.5	&	19:23:44.79	&	+14:29:11.2	&	J192344.79+142911.2	&	12.75 &	2015-07-01	&	1750	&	0.17	\\
W51e$_{1}$	&	O9	&	19:23:43.93	&	+14:30:27.8	&	J192343.93+143027.8	&	14.56 &	2015-07-03	&	1800	&	0.19	\\
W51e	&	9\,$\times$\,O4	&	19:23:42.28	&	+14:30:33.0	&	J192342.28+143033.0	&	11.36 &	2015-06-30	&	400	&	0.20	\\
W51e$_{2}$	&	O9.5	&	19:23:43.97	&	+14:30:32.1	&	J192343.97+143032.1	&	14.61 &	2015-07-02	&	1800	&	0.16	\\
W51g	&	O6	&	19:23:50.45	&	+14:32:57.4	&	J192350.45+143257.4	&	9.85 &	2015-06-30	&	50	&	0.23	\\
\enddata
\tablenotetext{a}{After revising the spectral types derived by \citet{meh94} from radio continuum flux according to the calibration tables presented by \citet{martins05}}.
\end{deluxetable*}
\setlength{\tabcolsep}{6pt}

The NIR counterparts were observed with the Near-Infrared Integral Field Spectrometer (NIFS) 
at the Gemini North Observatory. The data were obtained on 2015, May 01, 02 and 03, and July 30 (GN-2015A-Q-52, PI: C. Barbosa). NIFS provides three-dimensional imaging spectroscopy with spectral resolving power $R\sim5,200$ in the $K$-band (2.0-2.4\,\micron) over a field of view of {3\arcsec\,$\times$\,3\arcsec} \citep{mcg03}. The adaptive optics module ALTAIR was used during the observations in  laser guide star mode to correct the image of sources for the distortion of the Earth's atmosphere \citep{sad98}. The image quality of the AO corrected observations was estimated based on the full width at half maximum (FWHM) of an image extracted from the continuum. The typical spatial resolution of the images was obtained averaging the FWHM of each source, excluding those that exhibits extended emission (sources W51b$_2$ and W51e$_{7}$, see Sect. \ref{b2_map} and \ref{e7_map}) and W51e$_{2}$ due to the low S/N ratio of the image. The typical spatial resolution is about 0\farcs20 (0.27 pc at the new derived distance, see Sect. \ref{distance}).

The data were obtained in the $K$-band (NIFS grating `Z' and filter `HK') providing an effective spectral coverage from 1.99\,$\leq$\,$\lambda$\,$\leq$\,2.40\,{\micron} (central wavelength of $\lambda_{c}$\,=\,2.20\,\micron). For each on-source observation (position A), we took an associated image of a nearby blank sky (position B), offset by $\sim$10{\arcsec} from the target. This last image was used to subtract sky emission lines, mainly OH lines from airglow in the atmosphere, during the data reduction process.
As the OH emission lines vary in time scales as short as few minutes, the maximum individual exposure time was set to 400\,seconds. For the faint sources that required multiple exposures, we adopted an observation strategy between object and sky images following an ABAABA pattern. The observing dates, exposure time and achieved spatial resolution of the NIFS observations are listed in Table\,\ref{logs}. 

After an UC\ion{H}{2} region was observed, one spectrum of a Ne-Ar lamp was taken as a reference for wavelength calibration. A bright standard star (HIP\,95487, A0\,V) was observed right after each science target, matching the same airmass of the science observations. This star was used as a template to correct for the telluric absorption lines. The data reduction process is fully described in the next section.

\begin{figure}
\centering
\includegraphics{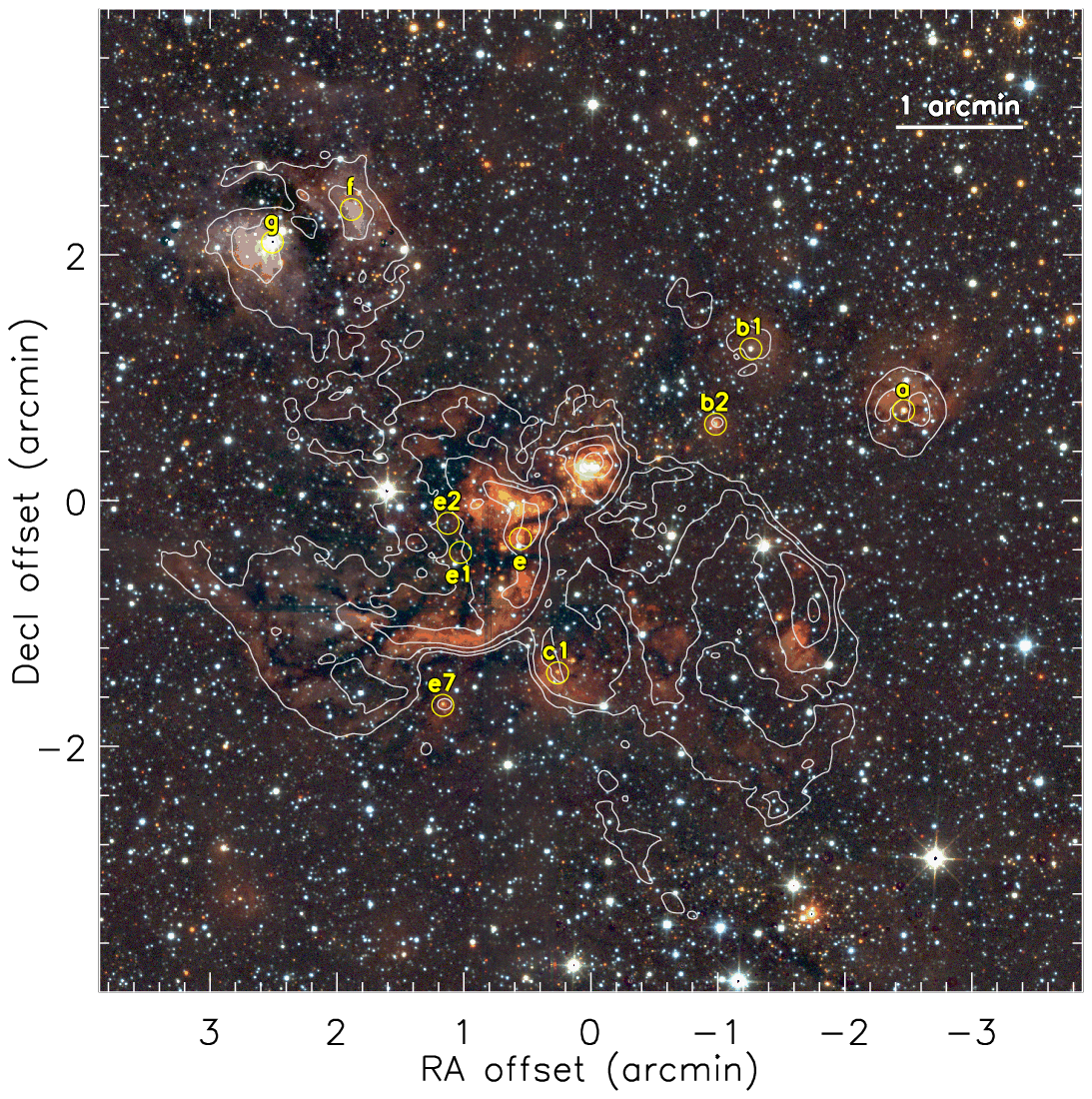}
\caption{False-color near-infrared image of the W51 North region. The image is color-coded as red ($K$-band), green ($H$-band) and blue ($J$-band) from the UKIDSS Survey \citep{law07}, overlaid by the 6 cm radio emission contours from \citet{meh94}. The contours are placed at 3$^n$-$\sigma$ level ($n$\,=\,1,2,3...). The UC\ion{H}{2} regions studied in this work are labeled and indicated by the yellow circles. The map corresponds to a field-of-view of {8\arcmin\,$\times$\,8\arcmin}, centered at RA\,=\,19:23:40.0, Dec\,=\,+14:30:51 (J2000)}. \label{fig_chart1}
\end{figure}

\subsection{Data reduction} \label{datared}
    
The standard data processing of NIFS observations was performed using the Gemini/NIFS \texttt{IRAF}\footnote{IRAF is distributed by the National Optical Astronomy Observatory, which is operated by the Association of Universities for Research in Astronomy, Inc., under cooperative agreement with the National Science Foundation.} package.
Calibration data were prepared with Gemini-\texttt{IRAF} \texttt{NFPREPARE} task which adds the Data Quality and Variance extensions to each FITS file.
    
A flat-field template was created using the set of flat-on and flat-off images. The wavelength transformation was performed using the Ne-Ar lamp spectrum. The RMS of the wavelength solution was $\leq0.15$\,{\AA} for each slit. The spatial rectification model was made using a Ronchi mask image \citep[for more information, see][]{blum08}.
    
The telluric and science observations were flat-fielded, wavelength corrected, and sky subtracted. The spectral template of the telluric lines was created by the extraction of a one-dimensional spectrum from a 0.5\arcsec\ radius aperture centered on the telluric standard star. Then, the stellar Br$\gamma$ feature was removed by fitting a Voigt profile between two continuum points. Next, the telluric correction was applied to both science and sky images using the \texttt{NFTELLURIC} task. 
    
Finally, the datacube was created by resampling the original rectangular spatial pixels to squared pixels of 0\farcs05 in both $x$- and $y$-directions. At the end of the processing, final datacubes with dimensions of $\sim$\,60\,$\times$\,60 spatial pixels and 2040 spectral pixels were delivered. A median-combined datacube was obtained from the input datacubes to suppress any bad pixels or possible cosmic rays present in the original data. 

\section{Results}
\label{results}

The one-dimensional spectra of the object were extracted performing a background subtraction of the spectrum of each source using a circular-annular aperture centered at the position of each object. Each spectrum was extracted through an inner radius of 4 pixels (0\farcs20) that was subtracted by the background spectrum obtained through an outer radius of 7 pixels (0\farcs35). After that, each spectrum was continuum-normalized by fitting a robust polynomial function of order $n$\,$\leq$\,4 on line-free regions of the spectrum. The results are presented in Figs.\,\ref{fig_normspec1}, \ref{fig_normspec2} and \ref{fig_normspec3}

We performed the spectral type classification of each source, first identifying photospheric lines using Peter van Hoof Atomic Line List\footnote{\url{https://www.pa.uky.edu/~peter/atomic/}} and then comparing their normalized spectra with the atlases presented by \citet{h96} and \citet{h05}. We detected photospheric features in seven of the nine objects in our sample. Only nebular features were identified in the spectrum of the W51e. We could detect no features in source W51e$_2$ due to its low signal to noise ratio spectrum, so its analysis was inconclusive. 

\subsection{Stellar spectral classification of the NIR counterparts of the UC\ion{H}{2}}
\subsection{Massive Stars}
\label{mass_st}

The continuum-normalized spectra of the NIR counterparts of the UC\ion{H}{2} are presented in Figs.\,\ref{fig_normspec1}, \ref{fig_normspec2} and \ref{fig_normspec3}. The spectral features used for classification of the spectral type of the sources are labeled on the plots. When compared to the spectral types derived from radio continuum data, our classification led to similar spectral types (within uncertainties), except for objects W51a and W51c$_{1}$. For these objects, we assigned a somewhat later spectral type. The extensive spectroscopy survey of UC\ion{H}{2} conducted by \citet{h02} found a similar trend, when a direct classification was possible through photospheric lines. The individual analysis and the spectral classification of each source is presented below. 

\subsubsection{W51a}
The spectrum of W51a is shown in Fig.\,\ref{fig_normspec1} (top). Several absorption features are observed in the spectrum of this object, including \ion{He}{1}\,(0--1) at 2.0581\,{\micron}, \ion{He}{1}\,(1--0) at 2.1137\,{\micron},
\ion{He}{2}\,(7--10) at 2.1885\,{\micron}, and \ion{H}{1}\,(4--7) at 2.1661\,{\micron} (Br$\gamma$). A broad \ion{N}{3}\,(7--10) emission is also observed at 2.1155\,{\micron}. \ion{N}{3} emission is seen in spectra of stars earlier than O9 and considering that \ion{C}{4} emission is not detected in spectra of stars later than O7, as the case of W51a, we can classify its ionizing source as an O8 star. Remaining emission lines in W51c$_{1}$ spectrum are artifacts introduced during the sky subtraction step.

\subsubsection{W51b$_{1}$}
The spectrum of W51b$_{1}$ is shown in Fig.\,\ref{fig_normspec1} (middle). This source exhibits the \ion{He}{1}\,(1--0) and Br$\gamma$ in absorption. These features suggest a spectral type between O9 -- B2. Based on the intensity of these transitions, specially the profile of \ion{He}{1} line, we narrowed the spectral classification of this object to an O9.5 star. Additional narrow features in emission in the blue region of the spectrum are residuals from the sky-subtraction procedure.

\subsubsection{W51b$_{2}$}
\begin{figure}
\centering
\scalebox{0.85}{\includegraphics{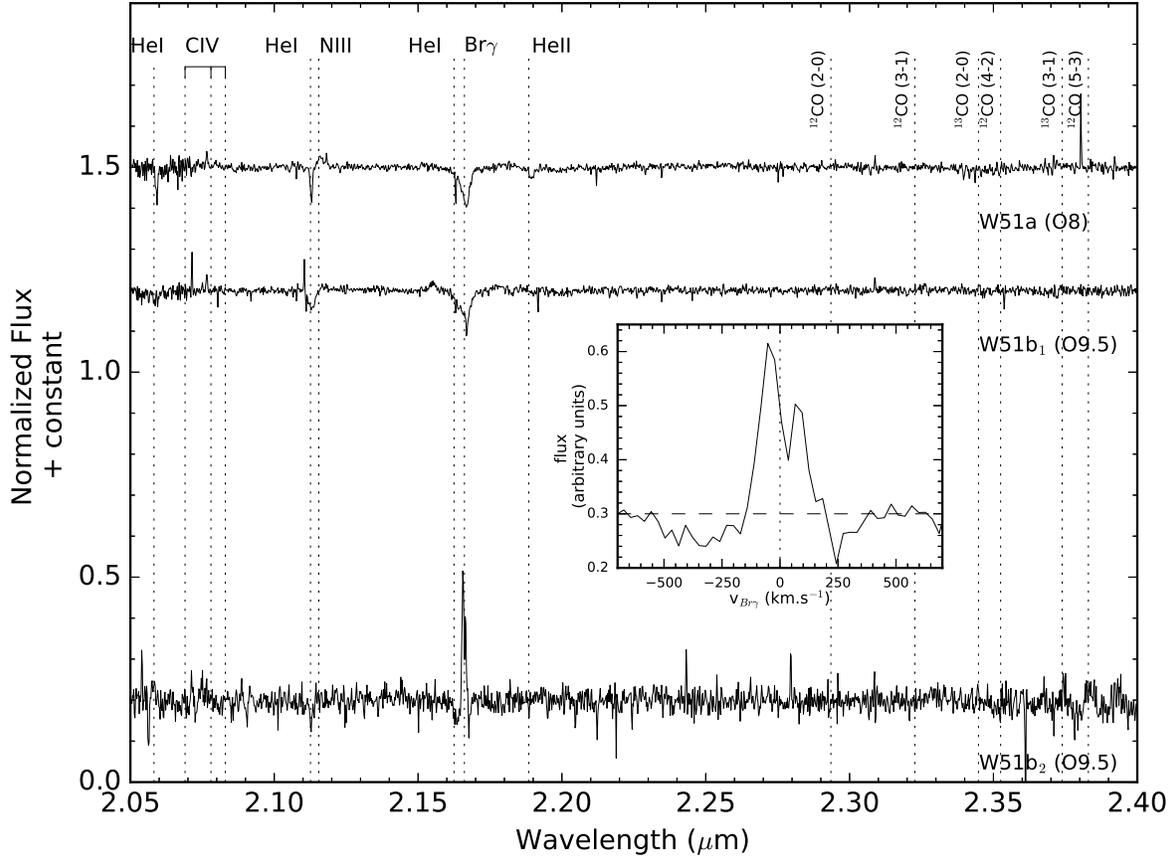}}
\caption{$K$-band spectra of the NIR counterparts of the UC\ion{H}{2} regions W51a (top spectrum), W51b$_{1}$ (middle) and W51b$_{2}$ (bottom). Based on the features identified in the spectra, we assigned the spectral type indicated in text. All spectra are continuum-normalized and were not corrected for the heliocentric velocity. The vertical dotted lines indicate typical photospheric lines used to determine the spectral type of massive stars, as much as the wavelength of $^{12}$CO and $^{13}$CO bandheads. The inset shows the double-peak profile of the Br$\gamma$ line observed in the spectrum of W51b$_{2}$. The horizontal axis is in velocity scale, centered at the vacuum rest wavelength of the line (vertical dotted line) and horizontal dashed line indicates the level of the continuum in the spectrum.}
\label{fig_normspec1}
\end{figure}

The spectrum of the NIR counterpart of W51b$_{2}$ (Fig.\,\ref{fig_normspec1} bottom) shows \ion{He}{1} (0--1) and Br$\gamma$ emission within a broad absorption profile. The analysis of the absorption features in the spectrum of W51b$_{2}$ suggests that it is a late O/early B star, however, the detection of a weak absorption profile of \ion{He}{2} (7--10) suggests that this object corresponds to an O9.5 star. Moreover, we see the emission component of the Br$\gamma$ feature is double peaked, as evident in the inset.

\subsubsection{W51c$_{1}$}
The spectrum of W51c$_{1}$ is presented in Fig.\,\ref{fig_normspec2} (top), showing the \ion{He}{1}\,(1-0) and the Br$\gamma$ features in absorption. These features are present in spectra of O9 -- B2 stars, however the \ion{He}{1}\,(0-1) is marginally detected, leading to a slightly earlier stellar type classification than that of W51b$_{1}$ (O9.5). Thus, we classified this object as an O9 star.

\subsubsection{W51e$_{7}$}
\begin{figure}
\centering
\scalebox{0.85}{\includegraphics{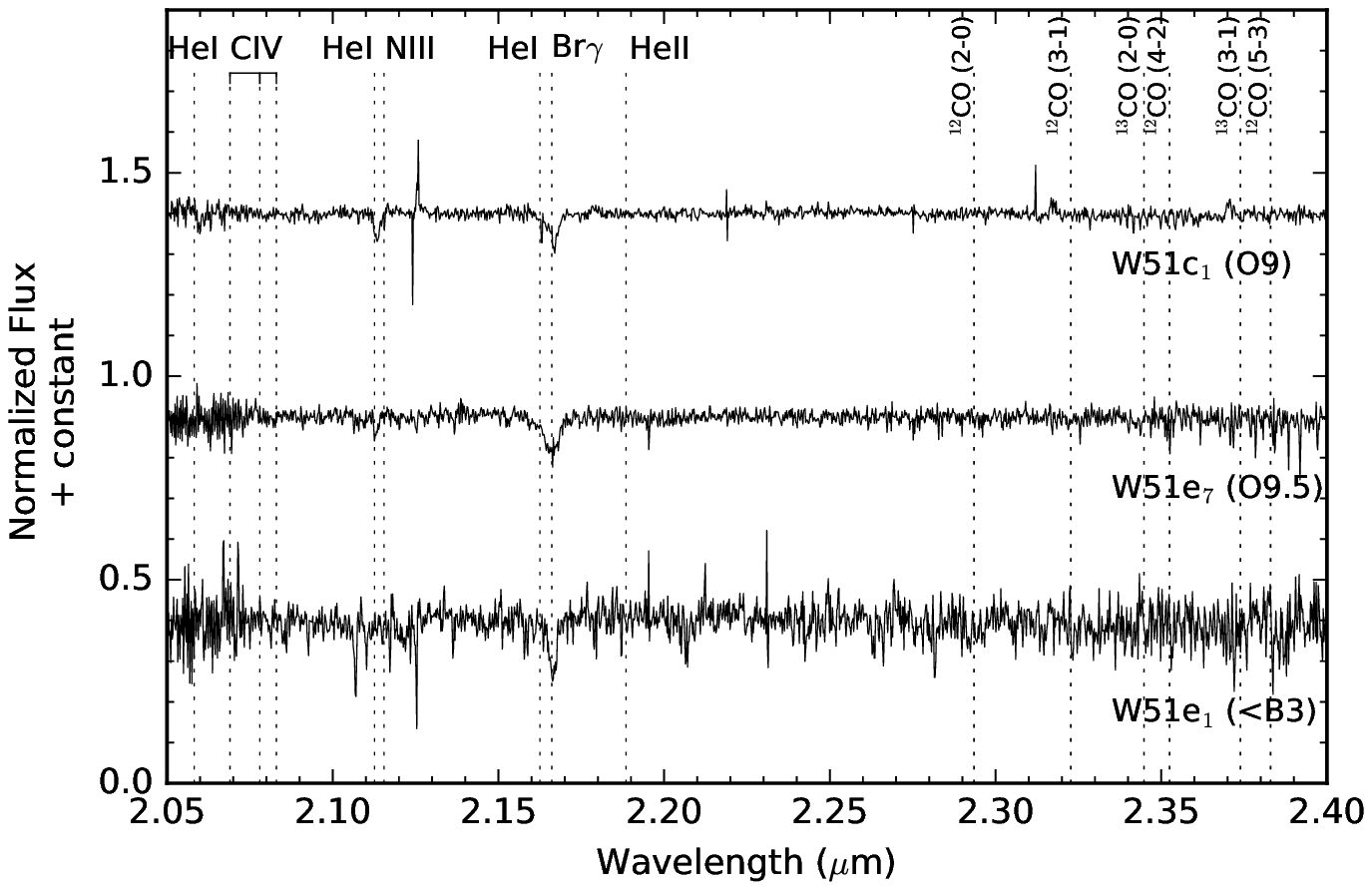}}
\caption{$K$-band spectra of the NIR counterparts of the UC\ion{H}{2} regions W51c$_{1}$, W51e$_{1}$ and W51e$_{7}$. Based on the features identified in the spectra, we assigned the spectral type indicated in the text. All spectra were normalized by their respective continuum and were not corrected for the heliocentric velocity. The vertical dashed lines indicate typical photospheric lines used to determine the spectral type of massive stars, as much as the wavelength of $^{12}$CO and $^{13}$CO bandheads. Emission lines in W51c$_{1}$ spectrum are artifacts introduced by the sky subtraction step.}
\label{fig_normspec2}
\end{figure}

The spectrum of source e$_{7}$ is presented in Fig.\,\ref{fig_normspec2} (middle), exhibiting the \ion{He}{1} in absorption and a relatively broad Br$\gamma$ absorption. The presence of these features suggests that the spectral classification is between O9--B2, but based on the intensity of both features, we classified this object as a O9.5 star. It is important to note that W51e$_{7}$ corresponds to object \#7 in the study of \citet{bik19} and they classified this object as a B1 star.

\subsubsection{W51e$_{1}$}
The spectrum of W51e$_{1}$ is shown in Fig.\,\ref{fig_normspec3} (middle). The only feature detected in the spectrum of W51e$_{1}$ corresponds to the Br$\gamma$ line in absorption, which solely is not sufficient to provide any constraint on the stellar type classification of the source. Since we identified this source as the NIR counterpart of an UCH\ion{H}{2} region, we assigned a lower limit to the spectral type of B3 for this object to account for ionizing radiation that produces the UCH\ion{H}{2} region. 

\subsubsection{W51e}
The background-subtracted spectra of source W51e does not show any photospheric lines, suggesting that it is still deeply embedded in its cocoon of gas and dust, at early stages of formation. Its spectrum is presented in Fig.\,\ref{fig_normspec3} (top). Emission lines in W51e spectrum are artifacts introduced during the sky subtraction step.

\subsubsection{W51e$_{2}$}
We present the spectrum of source W51e$_{2}$ in Fig.\,\ref{fig_normspec3} (middle). The spectrum does not show any spectral feature even after we smoothed it by a factor of 3 in an attempt to increase its signal to noise ratio, after the background subtraction. Thus, the nature of W51e$_{2}$ remains inconclusive.

\subsubsection{W51g}
\begin{figure}
\centering
\scalebox{0.85}{\includegraphics{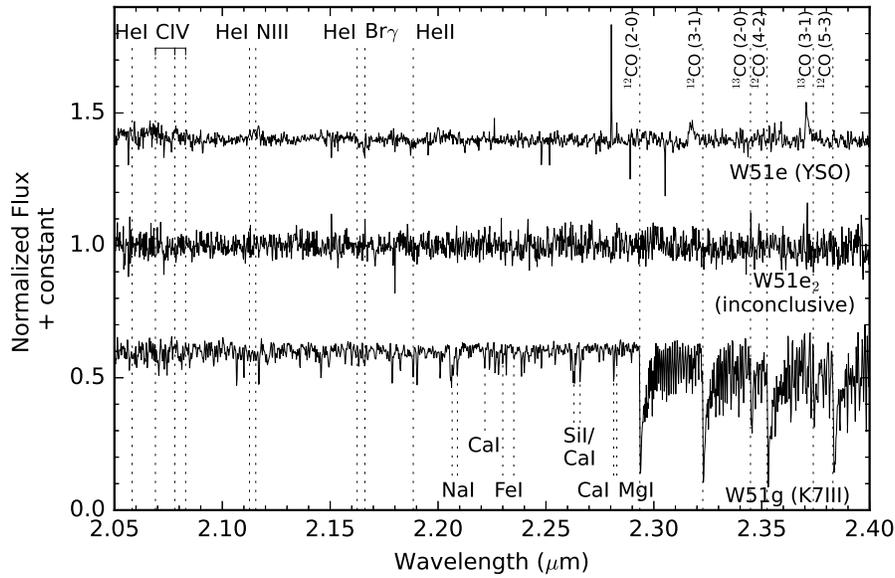}}
\caption{$K$-band spectra of the NIR counterpart of the UC\ion{H}{2} regions W51e (top), W51e$_{1}$ (middle) and W51g (bottom). The relatively low signal-to-noise spectrum of W51e$_{2}$ was smoothed by a factor of 3 in an attempt to identify any marginally detected photospheric line. The absence of photospheric features in the spectra of the W51e lead us to classify this source as a MYSO and W51e$_{2}$ remains inconclusive. The spectrum of the W51g exhibits only photospheric lines typical of late type stars, inconsistent with the NIR counterpart of a UC\ion{H}{2} region (see text for details). All spectra were normalized by their respective continuum and were not corrected for the heliocentric velocity. The vertical dotted lines indicate typical photospheric lines used to determine the spectral type of massive stars (top) and low mass stars (bottom), as much as the wavelength of $^{12}$CO and $^{13}$CO bandheads. Emission lines in W51e spectrum are artifacts introduced by the sky subtraction step.}
\label{fig_normspec3}
\end{figure}

The spectrum of the NIR counterpart of the UC\ion{H}{2} region W51g is presented in Fig.\,\ref{fig_normspec3} (bottom). Several photospheric absorption features were identified along the spectral range, such as the \ion{Na}{1} doublet at 2.2065 and 2.2089\,{\micron}, \ion{Ca}{1} lines at 2.2217 and 2.2814\,{\micron}, \ion{Ca}{1} and/or \ion{Si}{1} at 2.2658\,{\micron} and 2.2630\,{\micron}, \ion{Fe}{1} at 2.2351\,{\micron} and \ion{Mg}{1} at 2.2828\,{\micron}. Also, the first four $^{12}$CO bandheads were observed at 2.2935, 2.3227, 2.3525, and 2.3829\,{\micron}, as well as the $^{13}$CO bandheads at 2.3448 and 2.3739\,{\micron}. These features are not likely observed towards massive stars or MYSOs, corresponding to photospheric features typical of late type stars, and suggests that it is a K7 star. Therefore, the object identified as the NIR counterpart of W51g cannot be the ionizing source of the corresponding UC\ion{H}{2}.

 This object corresponds to source 4319847702260151680 in Gaia DR2 catalog \citep{gaia_dr2} and its distance is 2.11 kpc \citep{gaia_dist}. In fact, its continuum shows a negative slope towards longer wavelengths, indicating that this star is not subjected to the same extinction of the region and is a foreground star seen along the line of sight to W51g.

\subsection{Properties of the UC\ion{H}{2} Regions}

In the following subsections we present a brief review of each UC\ion{H}{2} region studied in this work, compiled from the literature.

\subsubsection{W51a}
W51a is resolved in a shell-like UC\ion{H}{2} region with a diameter of $\sim$1\,pc \citep{meh94}. Based on the timescale of the expansion of a Str\"omgrem sphere, \citet{oku00} estimated its age to be 0.7\,Myr, and based on their NIR photometry, they suggest that its NIR counterpart is actually a double MYSO of spectral types of O6+B1. Our results show that W51a is ionized by an O8 star. The difference may be explained by the environment surrounding the UC\ion{H}{2} that may impact the expansion rate of the Str\"omgrem sphere.

W51a has a remarkably different morphology at longer wavelengths, as recently shown by \citet{lim19}. Despite no point-like counterpart identified at $\lambda$\,$>$\,8\,{\micron}, the source exhibits an irregular multi-peaked morphology at 8\,{\micron}, and shows a round shape at 20\,{\micron} and 30\,{\micron}, brighter at the top and crossed by a dark lane on its equatorial region. 
 
\subsubsection{W51b$_{1}$}
The UC\ion{H}{2} region W51b$_{1}$ is not detected at $\lambda$\,=\,3.6\,cm, but has a diffuse morphology at $\lambda$\,=\,6\,cm, and exhibits a shell-like structure at $\lambda$\,=\,20\,cm, with a linear diameter of less than 1\,pc \citep{meh94}. \citet{oku00} classified the embedded source as an O9 star, and they estimated its age to be 0.8\,Myr based on the expansion of a Str\"omgrem sphere. Our results show that W51b$_{1}$ is ionized by an O9.5 star in good agreement.

$Spitzer$/IRAC images at 6 and 8\,{\micron} exhibit a point-like source and an arc-like feature surrounding it, with a diameter significantly smaller than that measured at 20\,cm (20\arcsec\ and 25\arcsec, respectively). This arc-like structure is not observed at $\lambda$\,=\,20\,{\micron}, but it clearly appears at $\lambda$\,=\,37\,{\micron}, giving a comet-like morphology to this source \citep{lim19}.

\subsubsection{W51b$_{2}$}\label{b2_map}
\citet{oku00} classified the ionizing source of W51b$_2$ as a B1 star, with an estimated age of 0.2\,Myr. Our results show that W51b$_{2}$ is ionized by an O9.5 star, also, in a good agreement. The \textit{Spitzer}/IRAC maps presented by \citet{barb16} exhibit W51b$_{2}$ off-centered from its surrounding MIR emission. SOFIA observations presented by \citet{lim19} exhibit this object as an unresolved point-like source at 20\,{\micron}, but extended at 37\,{\micron}.

\begin{figure}
\centering
\includegraphics[width=\linewidth]{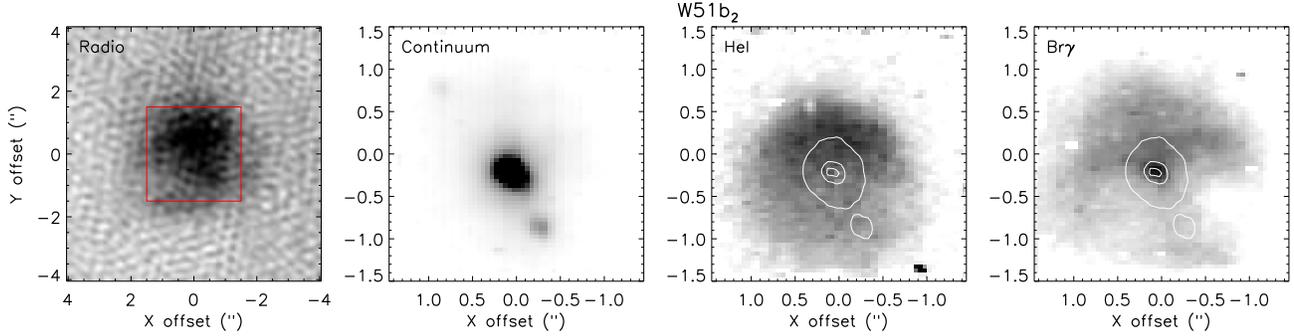}\\
\caption{Maps of the 6-cm continuum emission (left panel), NIR continuum, and the continuum-subtracted \ion{He}{1} (middle right) and Br$\gamma$ (right) features of source W51b$_{2}$. The white contours indicates the position of the point-like source, placed at 10, 50 and 90\% of the peak intensity of the continuum map. The red square on the radio continuum map represents the NIFS field of view. \label{b2_line_maps}}
\end{figure}

The UC\ion{H}{2} W51b$_{2}$ is a shell-like source at 6 and 20~cm, with no 3.6~cm counterpart \citep{meh94}. A similar inner \ion{He}{1} and outer Br$\gamma$ shell-like morphology is observed in the NIFS FOV, and we show the line maps of source b$_{2}$ in Fig.\,\ref{b2_line_maps}. The NIR continuum emission does not exhibit any extended emission associated with the NIR counterpart of the UC\ion{H}{2} region. A secondary point-like object is offset by $\sim$1\farcs0 to the SW of the main source; however, the signal is too weak to detect any photospheric features associated with it.

The Br$\gamma$ emission feature observed in W51b$_{2}$ (Fig. \ref{fig_normspec1}, inset) can be produced by either a disk \citep[e.g.,][]{blum04} or an expanding shell of gas \citep[see][and the references therein]{hart04}, as appears to be the case (see below). We measured its peak-to-peak velocity, $\Delta v$\,=\,116.6\,km\,s$^{-1}$ (FWHM of the Gaussian profile of 179.4\,km\,s$^{-1}$) and its velocity shift from rest wavelength $v$\,=\,(9.3\,$\pm$\,6.0)\,km\,s$^{-1}$, all values corrected to the heliocentric velocity $v_{\rm helio}$\,=\,6.5\,km\,s$^{-1}$.

Both the extended \ion{He}{1} and Br$\gamma$ morphologies are very similar to that observed in the radio continuum map at 6.2~cm, showing a shell-like structure with FWHM of $\sim$2\farcs.5, which seems to be the outer layer of the structures identified in the NIFS maps. The emission of the \ion{He}{1} transition exhibits a spherical structure showing stronger emission to the N of the NIR point-like source. The structure is nearly spherical with a FWHM of about 1\farcs7. The Br$\gamma$ emission exhibits a slightly different morphology, showing a ``mushroom-like" structure oriented in the N-S direction. The contrast of the Br$\gamma$ intensity is similar to that of the \ion{He}{1}, exhibiting a stronger emission to the N. The FWHM of the emission is larger than that probed by the \ion{He}{1}, 2\farcs1, suggesting the ionized hydrogen is located at a layer further out than the extended neutral helium emission.

\begin{figure}
\centering
\includegraphics[width=\linewidth]{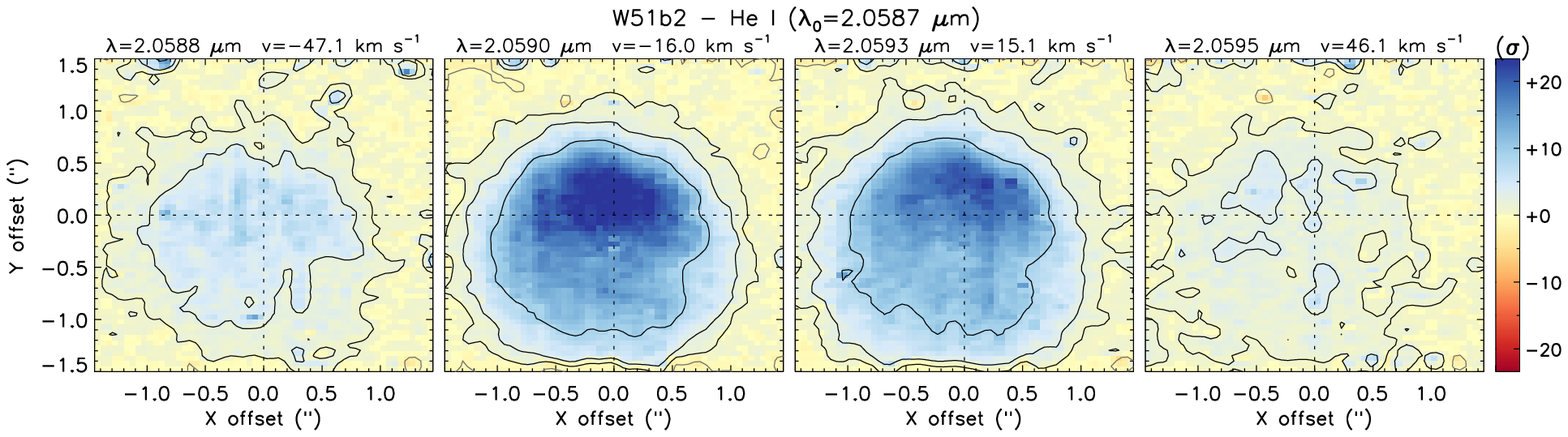}
\caption{W51b$_{2}$ velocity channel of \ion{He}{1} line. All velocity channels were corrected for the heliocentric velocity $v_{\rm helio}$\,=\,+6.54\,km\,s$^{-1}$ and $v_{\rm LSR}$\,=\,+60.0\,km\,s$^{-1}$. See Fig.\,\ref{e7_hei_map} for the full description of the plot. \label{b2_hei_map}}
\end{figure}

\begin{figure}
\centering
\includegraphics[width=\linewidth]{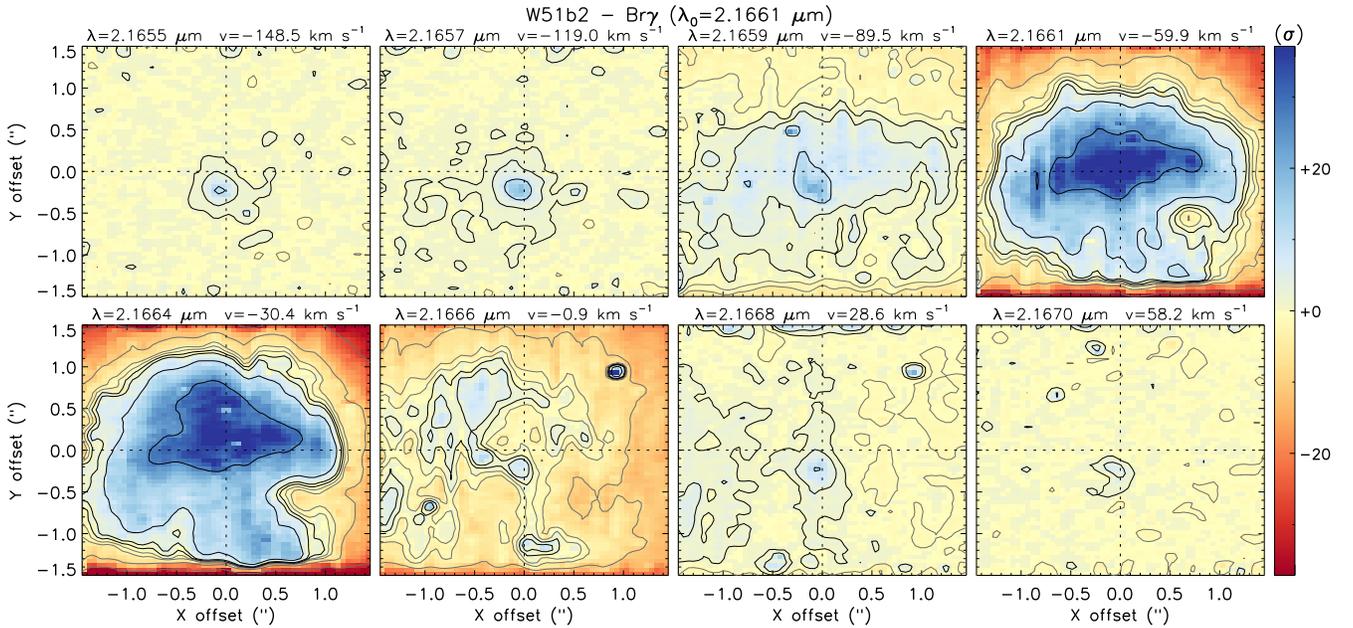}
\caption{W51b$_{2}$ velocity channel of the Br$\gamma$ line. All velocity channels were corrected for the heliocentric velocity $v_{\rm helio}$\,=\,+6.54\,km\,s$^{-1}$ and $v_{\rm LSR}$\,=\,+60.0\,km\,s$^{-1}$. See Fig.\,\ref{e7_hei_map} for the full description of the plot.\label{b2_brg_map}}
\end{figure}

Figures \ref{b2_hei_map} and \ref{b2_brg_map} present the velocity maps of the \ion{He}{1} and Br$\gamma$ lines, respectively for W51b$_2$. Both maps are compatible with the scenario derived from the emission line maps from Fig.\,\ref{b2_line_maps} in which an inner bubble of the slowly expanding shell of \ion{He}{1} is closer to the central star and is surrounded by a larger faster-expanding shell of ionized hydrogen.

\subsubsection{W51c$_{1}$}
Source W51c$_{1}$ appears as an arc-like UC\ion{H}{2} region surrounded by extended emission at radio wavelengths. The arc-like emission feature is also seen in the $Spitzer$/IRAC images reported by \citet{barb16}, and in 20\,{\micron} and 37\,{\micron} SOFIA maps from \citet{lim19}. A bow-shaped structure is more evident at 3.6\,cm \citep{meh94}.

\citet{oku00} estimated the age of this source as 0.4\,Myr. Based on their NIR photometry, these authors suggests that three sources (O5+O6+BO) are actually responsible for the ionization of the region, however, our results show that the ionizing object is an O9 star. The NIR counterpart of W51c$_{1}$ corresponds to source \#64 in the analysis presented by \citet{lys08}, but those authors did not specifically associate the object to the UC\ion{H}{2} region in their work.

\subsubsection{W51e$_{7}$}\label{e7_map}
Source W51e$_{7}$ was observed for the first time by \citet{meh94} who classified it as an UC\ion{H}{2}. According to its 6\,cm emission, this UC\ion{H}{2} is powered by a B1 star \citep{meh94,barb16}.  Its NIR counterpart was identified in the $K$-band image presented by \citet{lys08} and in the IRAC/\textit{Spitzer} images from \citet{barb16}. According to the NIR data presented by \citet{oku00}, source e$_{7}$ is a $\sim$\,0.3\,Myr B0 star. MIR images taken at 20\,{\micron} and 37\,{\micron} show e$_{7}$ as an unresolved source \citep{lim19}. These prior classifications based on emission characteristics can be compared to the somewhat hotter spectral classification spectral of O9.5V presented here.   

\begin{figure}
\centering
\includegraphics[width=\linewidth]{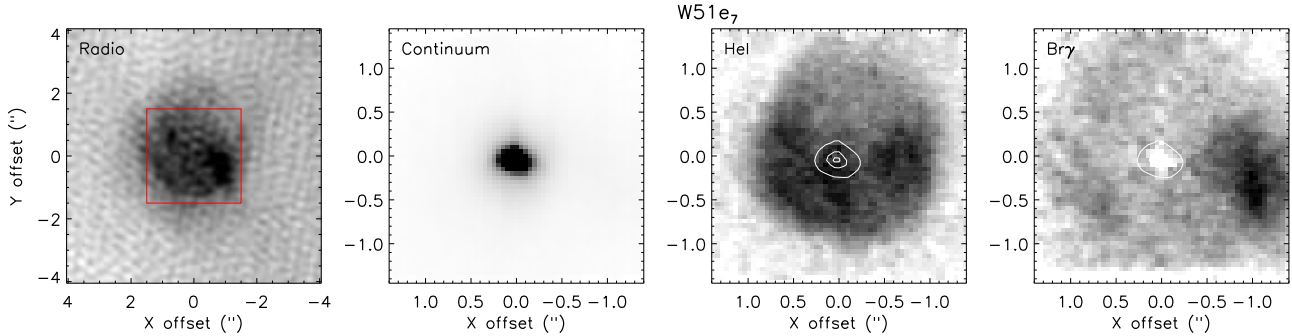} \\
\caption{Maps of the radio 6\;cm  continuum emission (left panel), NIR continuum (middle left), continuum-subtracted \ion{He}{1} (middle right) and Br$\gamma$ features (right) of source W51e$_{7}$. The white contours indicates the position of the point-like source, placed at 10, 50 and 90\% of the peak intensity of the continuum map. The red square on the radio continuum map represents the NIFS field of view. \label{e7_line_maps}}
\end{figure}

The NIFS observations were able to resolve the $K$-band emission in the inner {3\arcsec$\,\times$\,3\arcsec} region around the ionizing source of the UC\ion{H}{2} region. We present the emission line maps of W51e$_{7}$ in Fig.\,\ref{e7_line_maps}. The \ion{He}{1} emission exhibits a shell-like structure, with a FWHM of $\sim$\,2\farcs4 and 2\farcs2 in the major and minor axis, respectively. The Br$\gamma$ emission has a spherical shape with FWHM larger than $\sim$2\farcs5, fitting the entire FOV. 

It is interesting to note that, in spite of its shell morphology, the inner \ion{He}{1} emission bubble is more intense to the South, forming an arc shaped structure, which is also observed in the radio continuum map. The distribution of the ionized hydrogen gas probed by the Br$\gamma$ emission does not exhibit the arc-shaped structure, but it shows a bright spot to the West which also coincides with a peak in the radio emission. 

\begin{figure}
\centering
\includegraphics[width=\linewidth]{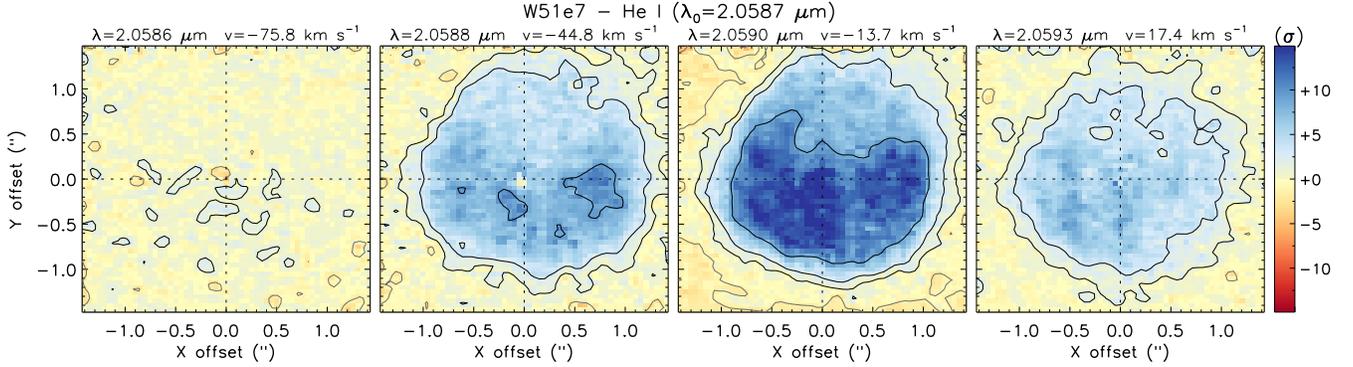}
\caption{W51e$_7$ velocity channel of \ion{He}{1} line. The corresponding wavelength and its relative velocity with respect to its rest wavelength are indicated in the top of each channel. The intensity is scaled in $\sigma$ units, where $\sigma$\,=\,rms, and is shown in a blue-to-red divergent scale, as indicated by the color bar at the right. The black contours are placed at $3^{(n - \sigma)}$, where $n$\,=\,0,\,1,\,2,\,$\dots$ and the grey contours are placed at ($-3^{n-\sigma}$) levels. All velocity channels were corrected for the heliocentric velocity $v_{\rm helio}$\,=\,+6.54\,km\,s$^{-1}$ and $v_{\rm LSR}$\,=\,+57.4\,km\,s$^{-1}$. \label{e7_hei_map}}
\end{figure}

\begin{figure}
\centering
\includegraphics[width=\linewidth]{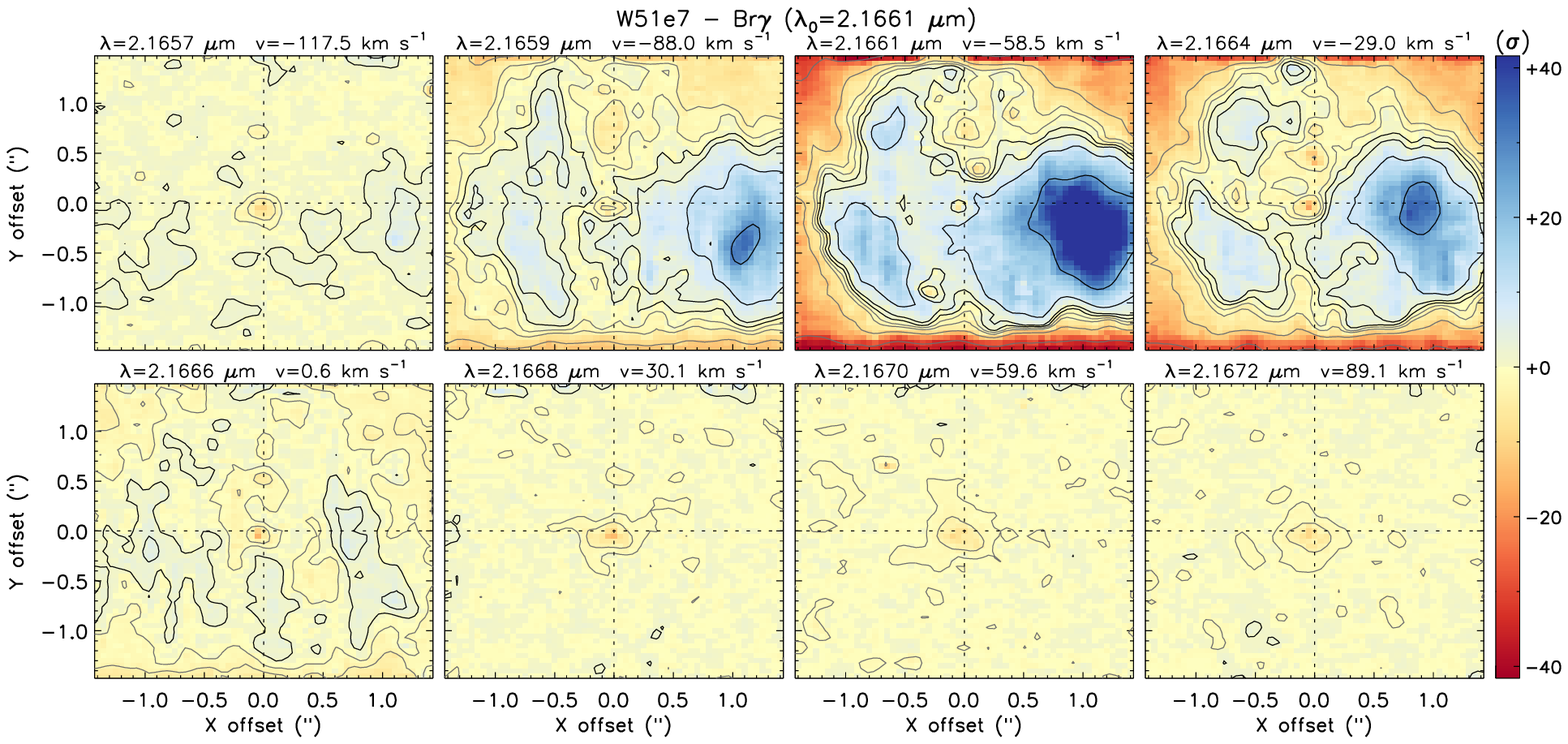}
\caption{W51e$_{7}$ velocity channel of the Br$\gamma$ line ($\lambda_{0}$\,=\,2.1661\,{\micron}). All velocity channels were corrected for the heliocentric velocity $v_{\rm helio}$\,=\,6.8\,km\,s$^{-1}$ and $v_{\rm LSR}$\,=\,+57.4\,km\,s$^{-1}$. See Fig.\,\ref{e7_hei_map} for the full description of the plot.\label{e7_brg_map}}
\end{figure}

The features revealed in the helium line map are also seen in the velocity maps of \ion{He}{1} and Br$\gamma$ presented in Figs.\,\ref{e7_hei_map} and \ref{e7_brg_map}, respectively. The \ion{He}{1} shell, centered at the NIR continuum source, is clearly seen at $v$\,=\,$-$13.7\,km.s$^{-1}$ and the arc shaped structure is evident at the $50$\,km\,s$^{-1}$ channel in Fig.\,\ref{e7_hei_map}. The channel maps of the Br$\gamma$ emission exhibits the bright spot to the W of the point-like object, but also indicate two fainter spots to the NE and SE of the source at $v$\,=\,$-$58.5\,km.s$^{-1}$, suggesting an inhomogeneous shell of ionized gas around the source.

\subsubsection{W51e$_{1}$}
Source W51e$_{1}$ exhibits a cometary morphology at radio wavelengths \citep{meh94} and it is also identified in the 4.5\,{\micron} \textit{Spitzer}/IRAC map presented by \citet{barb16}, offset by $\sim$2\arcsec\ of the radio peak. No point-like emission was identified in the NIR \citep{oku00}, in high-resolution ground-based MIR images \citep{barb16} and in MIR maps taken with SOFIA \citep{lim19}. Probably this is due to the intense emission from IRS1 \citep{barb16}.

\subsubsection{W51e}
Source W51e is a large arc-shaped UC\ion{H}{2} region with an extended radio emission spanning more than {50\arcsec}. This region exhibits the most intense radio emission among all the objects in W51A \citep{meh94} and based on the number of photons emitted in the Lyman continuum, \citet{meh94} proposed that the ionizing source is a cluster of nine O4 stars, or five O3 stars \citep[using the values for stellar fluxes presented by][]{martins05}. From NIR photometry, \citet{oku00} proposed a recipe (or a stellar equation) for the ionizing cluster: O4\,+\,O5\,+\,O6\,+\,O8\,+\,2$\times$B2. According to the latter authors, this cluster must not be older than 0.7\,Myr.

W51e is also the most luminous source of the whole complex in the infrared, which led \citet{wynn74} to identify this source as IRS1. As such, the NIR counterpart of W51e is surrounded by extended infrared emission, seen from 1.6\,{\micron} to longer wavelengths. In the MIR, this source is unresolved in maps obtained at 7, 12.3 and 24.5\,{\micron} \citep{barb16}, and 20 and 37\,{\micron} \citep{lim19}.
The NIR counterpart of this UC\ion{H}{2} corresponds to source \#45 in the study presented by \citet{lys08}. It is interesting to note that  the source \#44 -- one of the four O-type stars identified by those authors (O5) in the main cluster -- is offset by less than {5\arcsec} to the South of W51e.

\subsubsection{W51e$_{2}$}
Sources W51e$_{1}$ and W51e$_{2}$ form a subcluster within the extended radio emission arising from W51e. Both sources are likely young embedded O-type stars and, together with IRS2, represent the main regions of ongoing star formation activity within W51A \citep{ginsb15}.
 
The NIR counterpart of the HC\ion{H}{2} region W51$e_{2}$ was not identified by \citet{oku00}, but within the 3\arcsec\ search radius, a point-like object can be identified in the $K$-band images from \citet{lys08} (source \#145, about $\sim$1\arcsec\ from the e$_{2}$ radio peak). High-resolution ground-based MIR observations taken with T-ReCS were unable to identify any emission towards W51e$_{2}$ \citep{barb16}, while this object appears as an unresolved emission in SOFIA maps at longer wavelengths \citep{lim19}.
 
\subsubsection{W51g}
Source W51g has a core-halo morphology in the scenario proposed by \citet{wc89a}, and its ionizing source is an O6 star \citep{meh94}. It is interesting to note that sources W51g and W51f form two extended lobes, almost symmetrical in both 6- and 20-cm radio maps, but they were not detected at 3.5-cm \citep{meh94}. No NIR counterpart was found between the lobes, nor near the peak of the radio emission of source W51f.

Instead, there is a source $\sim${6\arcsec} west of the peak of the W51g radio emission.
The offset is larger than the search radius defined in Sect.\,\ref{target_selection} ({3\arcsec}). nevertheless, this target was chosen as the best NIR counterpart candidate for the UC\ion{H}{2}. As noted above it turns out to be a foreground late--type star. Thus no candidate NIR counterpart has been identified for either W51g or W51f. \textit{Spitzer}/IRAC images presented by \citet{barb16} show both UC\ion{H}{2} regions W51g and W51f permeated by the extended green emission at 4.5\,{\micron} that may correspond to the ionized gas traced by the Br$\alpha$ emission at 4.09\,{\micron}. The intense emission appears to be hiding the NIR sources.

\section{The spectrophotometric distance to W51A}
\label{distance}
 
The spectral classification of the stellar objects associated with the UC\ion{H}{2} regions was used to estimate their spectrophotometric distance and, hence, the distance to the main cluster itself.
In addition, we included the objects \#44, \#50, \#57 and \#61 from \citet{lys08} and sources \#6, \#12 and \#13 from \citet{bik19}, as these sources are also members of W51A. We used the spectral types assigned by those authors assuming all of them as dwarfs (luminosity class V). 

The calculation of a spectrophotometric distance $d$ is a simple procedure that involves the classification of the objects by spectral type,  the assumption of a reddening law, and the inversion of the distance modulus equation:
\begin{equation}
    m_{K} - M_{K} = 5 \cdot \log(d) - 5 + A_{K}
\label{dist_mod}
\end{equation}
\noindent where $m_{K}$ and $M_{K}$ are, respectively, the apparent and absolute $K$-band magnitudes of the object, $A_{K}$ is the extinction in the line of sight of the source at the same wavelength, and d is the distance in parsec.

For the spectral type assigned to each object, we used the corresponding effective temperature, absolute magnitudes and bolometric corrections for O-type stars reported by \citet{martins06}. For B-type stars, we obtained the absolute magnitudes extrapolating the relation between visual absolute magnitudes and spectral types for O stars presented by \citet{martins05}. Then, $K-$band absolute magnitudes were obtained using intrinsic colors from \citet{pec13}, as much as their effective temperatures and bolometric corrections. In order to use an homogeneous data set, we obtained the NIR photometry and the color indexes from the UKIDSS Galactic Plane Survey DR6 \citep{luc08}. We adopted the extinction law presented by \citet{dami16} (hereafter D16). The amount of extinction was estimated averaging $A_{K}$ from its relation with the color excess obtained from the $JHK_{S}$ colors as:
\begin{eqnarray}
A_{K} & = & 0.449\cdot(E_{J-K})\\
A_{K} & = & 0.685\cdot(E_{J-H})\\
A_{K} & = & 1.30\cdot(E_{H-K}) 
\end{eqnarray}

The color excess of each NIR counterpart was calculated using the colors listed in Table\,\ref{logs}, and adopting the intrinsic colors from \citet{martins06} and from \citet{pec13}, for O- and B-type stars, respectively. 

The results are presented in Table\,\ref{tab_dist}. Distances considering only individual sources range between 2.83$_{-0.49}^{+0.59}$ and 6.44$_{-1.11}^{+1.34}$\,kpc. The uncertainty in the individual distances reflects the intrinsic scatter of $\pm$\,0.41\,mag in the $M_{V}$ scale presented by \citet{martins05} for the O-type stars. In the absence of such information for B-type stars, we adopted the same scattering for the $M_{V}$ values of these objects. Finally, we obtained a distance of $d$\,=\,4.80\,$\pm$\,1.27\,kpc after averaging the distances to each source of the cluster.

\begin{deluxetable}{lcccccccc}
\tablecaption{Physical parameters of objects in W51A and its spectrophotometric distance\label{tab_dist}}
\tablecolumns{8}
\tablewidth{0pt}
\tablehead{
\colhead{Object} & \colhead{Sp. Ty.} & \colhead{$K_{S}$} & \colhead{$(J-H)$} & \colhead{$(H-K_{S})$} & \colhead{A$_{K}$} & \colhead{$\log{(T_{eff})}$} & \colhead{$\log{(L_{bol}/L_{\sun})}$} & \colhead{$d$}\\
\colhead{} & \colhead{NIR} & \colhead{(mag)} & \colhead{(mag)} & \colhead{(mag)} & \colhead{(mag)} & \colhead{(K)} & \colhead{} & \colhead{(kpc)}
}
\startdata
W51a & O8 & 10.930\,$\pm$\,0.001 & 2.432\,$\pm$\,0.003 & 1.354\,$\pm$\,0.001 & 1.71\,$\pm$\,0.03 & 4.54\,$\pm$\,0.01 & 5.25$_{-0.27}^{+0.20}$ & 3.58$_{-0.62}^{+0.74}$\\
W51b$_{1}$ & O9.5 & 11.880\,$\pm$\,0.001 & 1.496\,$\pm$\,0.002 & 0.858\,$\pm$\,0.001 & 1.07\,$\pm$\,0.03 & 4.50\,$\pm$\,0.02 & 4.50$_{-0.27}^{+0.20}$ & 6.17$_{-1.06}^{+1.28}$\\
W51b$_{2}$ & O9.5 & 12.731\,$\pm$\,0.002 & 2.303\,$\pm$\,0.016 & 1.978\,$\pm$\,0.004 & 2.02\,$\pm$\,0.36 & 4.50\,$\pm$\,0.02 & 4.54$_{-0.27}^{+0.20}$ & 5.89$_{-1.01}^{+1.22}$\\
W51c$_{1}$ & O9   & 12.691\,$\pm$\,0.002 & 3.779\,$\pm$\,0.044 & 1.755\,$\pm$\,0.004 & 2.45\,$\pm$\,0.11 & 4.52\,$\pm$\,0.03 & 4.77$_{-0.27}^{+0.20}$ &  5.06$_{-0.87}^{+1.05}$\\
W51e$_{7}$ & O9.5 & 13.093\,$\pm$\,0.003 & 2.915\,$\pm$\,0.034 & 1.874\,$\pm$\,0.006 & 2.19\,$\pm$\,0.16 & 4.50\,$\pm$\,0.05 & 4.47$_{-0.27}^{+0.20}$ & 6.44$_{-1.11}^{+1.34}$\\ \hline
\#44  & O5   & 11.123\,$\pm$\,0.001 & 4.676\,$\pm$\,0.049 & 2.546\,$\pm$\,0.002 &  3.25\,$\pm$\,0.04 & 4.61\,$\pm$\,0.02 & 6.07$_{-0.27}^{+0.20}$ & 2.83$_{-0.49}^{+0.59}$\\
\#50  & O6.5 & 12.242\,$\pm$\,0.001 & 3.483\,$\pm$\,0.030 & 1.981\,$\pm$\,0.003 & 2.47\,$\pm$\,0.07 & 4.58\,$\pm$\,0.02 & 5.16$_{-0.27}^{+0.20}$ & 5.65$_{-0.97}^{+1.18}$\\
\#57  & O4   & 10.762\,$\pm$\,0.001 & 3.926\,$\pm$\,0.013 & 2.175\,$\pm$\,0.001 & 2.75\,$\pm$\,0.05 & 4.63\,$\pm$\,0.02 & 6.07$_{-0.27}^{+0.20}$ & 3.45$_{-0.59}^{+0.72}$\\
\#61  & O7.5 & 12.412\,$\pm$\,0.003 & \nodata & 2.823\,$\pm$\,0.030 & 3.67\,$\pm$\,0.06 & 4.55\,$\pm$\,0.01 & 5.49$_{-0.25}^{+0.18}$ & 3.10$_{-0.53}^{+0.64}$\\ \hline
\#6 & B1 & 12.691\,$\pm$\,0.002 & 3.779\,$\pm$\,0.044 & 1.755\,$\pm$\,0.004 & 2.45\,$\pm$\,0.11 & 4.43\,$\pm$\,0.10 & 4.66$_{-0.27}^{+0.20}$ & 3.87$_{-0.67}^{+0.81}$\\
\#12 & B2 & 13.783\,$\pm$\,0.005 & 2.140\,$\pm$\,0.079 & 2.865\,$\pm$\,0.022 & 2.48\,$\pm$\,0.81 & 4.30\,$\pm$\,0.10 & 3.91$_{-0.27}^{+0.20}$ & 5.50$_{-0.95}^{+1.14}$\\
\#13 & B5 & 13.766\,$\pm$\,0.004 & 1.780\,$\pm$\,0.014 & 1.337\,$\pm$\,0.006 &  1.45\,$\pm$\,0.19 & 4.19\,$\pm$\,0.04 & 3.15$_{-0.27}^{+0.20}$ & 6.07$_{-1.04}^{+1.26}$\\ \hline
 & & & & & & & \textbf{Average} & 4.80 $\pm$ 1.27\\
\enddata
\tablecomments{Column 1 gives the name of the object. Column 2 gives the spectral type assigned from $K$-band spectra. Column 3 gives the apparent magnitude in the $K_{S}$ band. Column 4 gives the $J-H$ color of each object. Columns 5 gives the $H-K_{S}$ color. Columns 6 lists the amount of the extinction in the $K_{S}$ band, after \citet{dami16}. Column 7 gives the effective temperatures taken from \citet{martins05} and \citet{pec13} for O- and B-type stars, respectively. The errors in the temperature represent an uncertainty of one spectral subtype. Column 8 gives the bolometric luminosity, the uncertainty quoted in the luminosity represents its variation based on the uncertainty in the distance to the cluster. Column 9 presents the spectrophotometric distance for each object, calculated using Eq.\,\eqref{dist_mod}, and its uncertainty represents the variation in distance due to the uncertainty in the extinction quoted in Column 6. The last row of the table presents the average distance to the W51A and its standard deviation.}
\end{deluxetable}

\subsection{The new spectrophotometric distance to W51 Main in context}
\label{dist}

\begin{figure}
\centering
\includegraphics{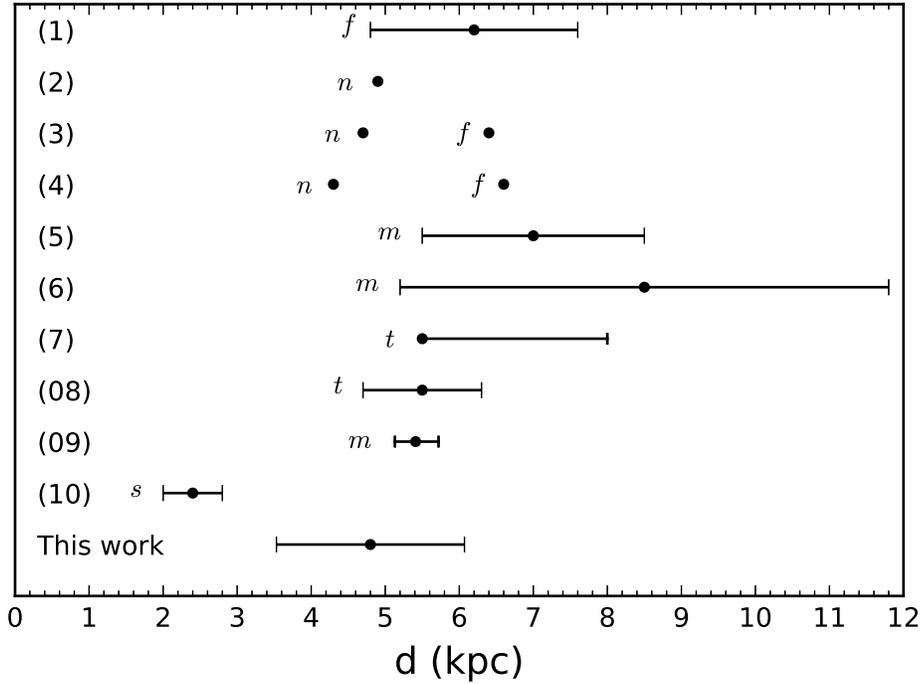}
\caption{The distance to W51 Main derived by different works in the literature: $n$ and $f$ represents the near and far kinematic distance, respectively;
$t$ indicates that the distance was obtained assuming that the cluster is at the tangent point of the corresponding spiral arm,
$m$ indicates a maser distance, and
$s$ is for spectrophotometric distance. References: (1) \citet{wilson70}, (2) \citet{georg76}, (3) \citet{cramp78}, (4) \citet{down80}, (5) \citet{genzel82}, (6) \citet{ima02}, (7) \citet{rus03}, (8) \citet{kolp03}, (9) \citet{sato10} and (10) \citet{lys08}.} \label{d_fig}
\end{figure}

We compare our spectrophotometric distance to W51 with the previous previous distances to the W51 complex available in the literature summarized in Table\,\ref{d_comp}, plotting them in Fig.\,\ref{d_fig}. The figure shows that the spectrophotometric distance derived in this work is compatible, within the uncertainties, with the near kinematic distances from \citet{georg76,cramp78,down80} and also the distances obtained by trigonometric parallaxes of water masers from \citet{genzel82,ima02,sato10}.

\citet{lys08} reported the spectrophotometric distance of $d$\,=\,2.0\,$\pm$\,0.3\,kpc, based on the analysis of 4 NIR objects associated with W51, \#44, \#50, \#57 and \#61, assuming that all are still on the zero age main sequence (ZAMS). This distance is substantially smaller than the one obtained in this work and we further discuss a possible explanation for that in Section \ref{discuss}.

\begin{deluxetable}{lll}
\tablecaption{Distance to W51\,Main from previous works in the literature. \label{d_comp}}
\tablecolumns{3}
\tablewidth{0pt}
\tablehead{
\colhead{$d$ (kpc)} & \colhead{Method} & \colhead{Reference}
}
\startdata
6.2\,$\pm$\,1.4                    & kinematic distance\tablenotemark{(a)} & (1) \citet{wilson70} \\
4.9\tablenotemark{(b)}             & kinematic distance\tablenotemark{(a)} & (2) \citet{georg76} \\
4.7/6.4\tablenotemark{(b)}         & kinematic distance\tablenotemark{(a)} & (3) \citet{cramp78}\\
4.3/6.6\tablenotemark{(b)}         & kinematic distance\tablenotemark{(a)} & (4) \citet{down80}\\
7.0\,$\pm$\,1.5                    & water masers & (5) \citet{genzel82}\\
8.5\,$\pm$\,3.3                    & water masers & (6) \citet{ima02}\\
5.5$_{-6.1}^{8.0}$\tablenotemark{(c)} & kinematic distance & (7) \citet{rus03}\\
5.5\,$\pm$\,0.8                    & kinematic distance    & (8) \citet{kolp03}\\
5.41$^{+0.31}_{-0.28}$             & water masers          & (9) \citet{sato10}\\
2.4\,$\pm$\,0.4\tablenotemark{(d)} & NIR spectrophotometry & (10) \citet{lys08}\\
4.80\,$\pm$\,1.27                  & NIR spectrophotometry & This work\\
\enddata
\tablenotetext{(a)}{\;\; Corrected for $R_{0}$\,=\,8.5\,kpc}
\tablenotetext{(b)}{\;\; No errors quoted}
\tablenotetext{(c)}{\;\; Limits to the distance obtained when the systemic velocity varies within the systematic error of $\pm$ 5 km s$^{-1}$.}
\tablenotetext{(d)}{\;\; Distance considering objects as Main Sequence stars.}
\end{deluxetable}

\subsection{HR diagrams}\label{HRD}

\begin{figure*}
\centering
\includegraphics[width=0.45\linewidth]{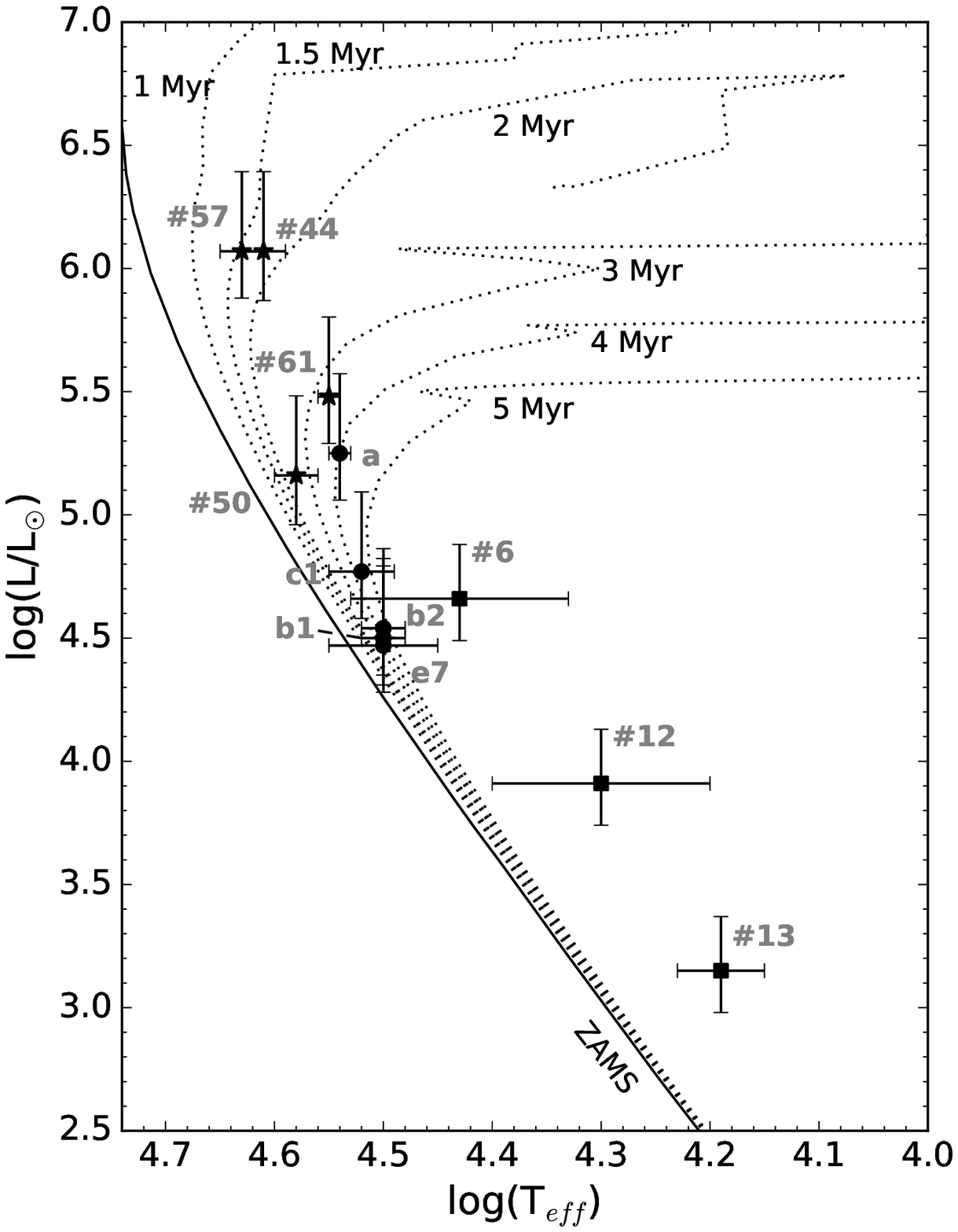}
\includegraphics[width=0.45\linewidth]{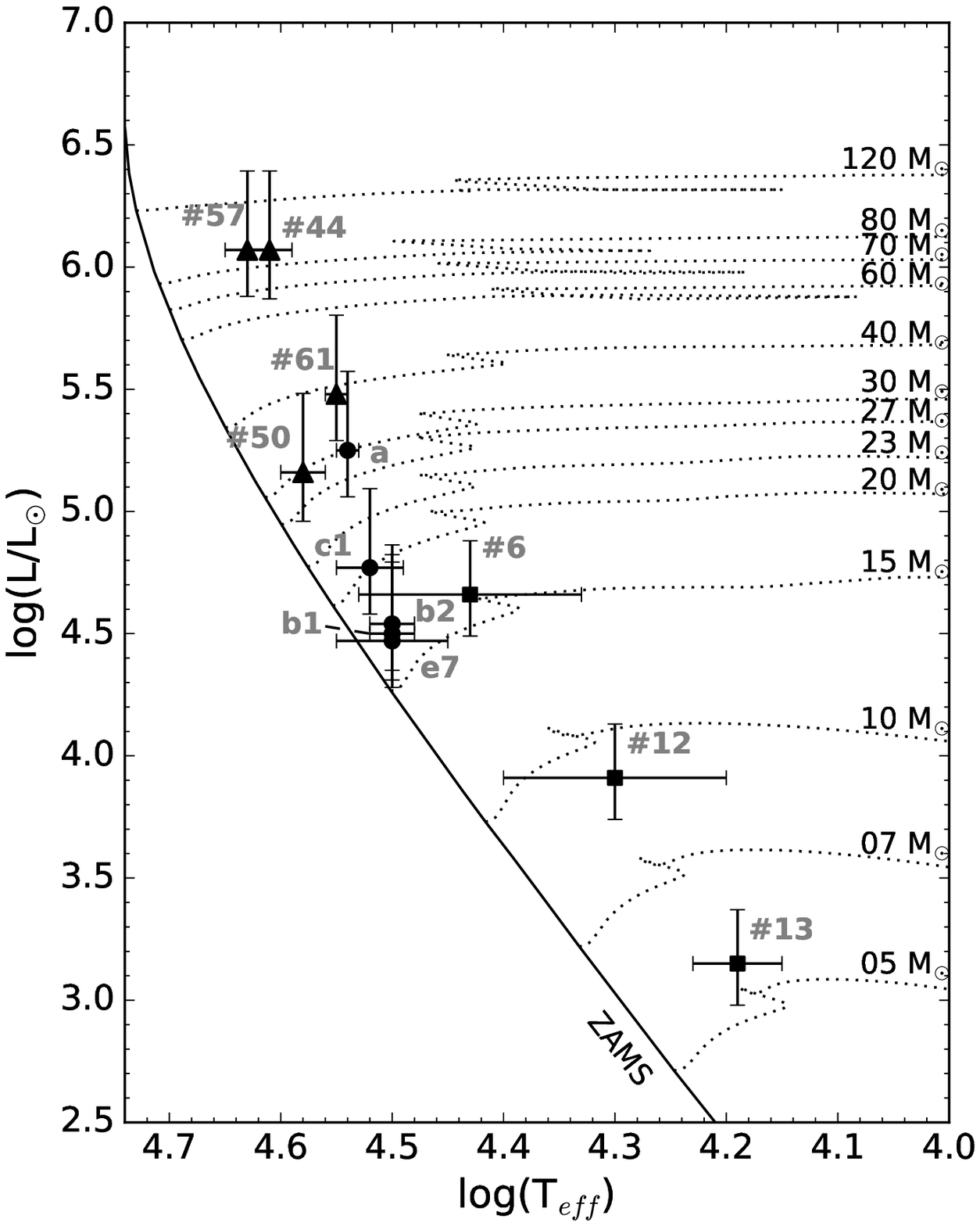}
\caption{Hertzsprung--Russell (HR) diagram of the UC\ion{H}{2} regions studied in this work (black circles), those presented by \citet{lys08} (black triangles) and those presented by \citet{bik14} (black squares). The photometric data were corrected using the extinction law from \citet{dami16}. In both panels, the solid line represents the ZAMS isochrone and the dotted lines are the main-sequence isochrones as a function of age (left panel) and initial mass (right), derived from the evolutionary tracks (including rotation and solar metallicity) presented by \citet{eks12}. \label{hr_d16}}
\end{figure*}

Using the distance to W51A, we constructed the Hertzsprung-Russell (HR) diagram of the cluster, considering the extinction law \citet{dami16}. We estimated the bolometric luminosity of the sources in W51A as follows. First, we corrected the apparent magnitudes for the extinction and then converted them into absolute magnitudes scaling for the distance $d$ to the cluster presented in Table\,\ref{tab_dist}. After that, we applied the bolometric correction from \citet{martins06} (for O-type stars) and \citet{pec13} (for B-type stars). Finally, we calculated the bolometric luminosities from the absolute magnitudes. We used the bolometric luminosities and effective temperatures from Table\,\ref{tab_dist} to construct the HR diagram presented in Fig.\,\ref{hr_d16}.

The uncertainty of the effective temperature represents the difference between the effective temperature of the assigned spectral type and the temperature of its subsequent subtype both earlier and later, leading to asymmetric upper and lower errors in most of the cases. For these sources, we considered the larger error value.

The main source of uncertainty of the bolometric luminosity of the sources arises from the uncertainty on the distance to the cluster. For this reason, the error bars in luminosity reflect the luminosities calculated taking account of the quoted uncertainty in the distance to the cluster, i. e., the upper limit corresponds to the luminosity considering the distance $d + \sigma_{d}$, and the lower limit was obtained using $d - \sigma_{d}$.

\section{Discussion}
\label{discuss}

In this section, we discuss the results presented in the previous sections, putting them in context, and comparing them with those available in literature.

Spectrophotometric distances are critically dependent on the assumed extinction law. As such, we obtained the distance to W51\,Main cluster as $d$\,=\,4.80\,$\pm$\,1.27\,kpc, including the objects previously identified by \citet{lys08} (after changing luminosity class from ZAMS to Main Sequence) and \citet{bik14} and adopting the extinction law from \citet{dami16}. Within the errors, this result is compatible with kinematic distances and distances obtained through maser parallaxes, also reconciling with previous studies that located the region near to the tangent point of the Carina-Sagittarius Arm \citep{kolp03}.

\citet{lys08} obtained a distance to W51A of 2.40\,$\pm$\,0.4\,kpc (assuming the objects as class V stars). The difference between their results and the results presented in this work is due to the extinction law adopted by them, \citet{mat90}: $A_{K}$\,=\,1.7$\times$\,$(E_{H-K})$. The adoption of the \cite{dami16} law results in an average extinction at $K$ of 2.3 mag compared to 3.9 mag for \cite{lys08} for the four stars 44, 50, 57, and 61. 

A distance of about 2.50\,kpc for W51A seems unlikely and \cite{lys08} concluded there was no satisfactory explanation for the short distance compared to the longer kinematic and maser distances. This distance is closer than the near solution for the kinematic distance of W51A \citep[4.3\,kpc,][]{down80}. Moreover, the line-of-sight of Galactic longitudes around $\ell$\,=\,49.5$^{\circ}$ seems devoid of Giant \ion{H}{2} regions for distances shorter than $\sim$4\,kpc \citep{rus03}. In retrospect, the extinction law and evolved nature of the stars nicely explain the difference. Indeed our results suggest that any O star that can be observed to exhibit absorption lines is necessarily evolved from the ZAMS. This means the timescale to clear itself from the overlying material related to its formation is similar to the time to evolve from the ZAMS for more massive stars.

The distance to W51A, calculated in Sect.\,\ref{dist}, allowed us to further investigate the nature of the sources presented in this work. The HR diagrams presented in Fig.\,\ref{hr_d16} are consistent with a young cluster still forming stars, placing sources \#57 and \#44 as the youngest (1.5-2.0\,Myr) and as the most massive objects ($\geq$\,80\,M$_{\sun}$) in W51A.

Objects \#61, \#50 and W51a form a group of massive stars with masses ranging between 30\,$\leq$\,M\,$\leq$\,40\,M$_{\sun}$ and ages formally between 3-4\,Myr, but \#50 is consistent with an younger age. Stars \#60 and W51a are offset from the center of W51A Main/IRS1 and may indeed represent a slightly earlier epoch of star formation. A third group of objects with masses ranging between 15-20\,M$_{\sun}$ and ages between 3-6\,Myr is formed by objects W51b$_1$, W51b$_2$, W51c$_1$, W51e$_7$ and \#6. The lowest mass end of the objects studied in this work is constituted by object \#12, with mass $leq 10$. Source \#13 may represent a intermediate mass ($\sim$\,5\,M$_{\sun}$) cluster member. The ages of these objects are not constrained by our observations.

The HR diagrams presented in Fig.\,\ref{hr_d16} differ from those presented by \citet{bik19} mainly because of the choice of the distance to the cluster and the reddening law adopted in their analysis. While we adopted 4.80\,kpc and the reddening law of \citet{dami16}, \citet{bik19} used 5.39\,kpc (distance modulus of 13.66) and adopted \citet{nishi09} for the extinction law.

Previous authors suggested that W51A is a very young cluster, with age within 0.5--1 Myr \citep[e.g.,][]{gww94,oku00,oku01}. However, recent multi-wavelength studies \citep[radio][]{ginsb15}, \citep[MIR][]{lim19}, \citep[NIR][]{bik19} and this work, show that this cluster must be older. It is very difficult to assess the star formation history within such a large molecular cloud, like the one W51A is embedded in. \citet{kumar04}, suggests that, due to its location near the tangential point of the Carina-Sagittarius arm, density waves may have started the formation of stars in the cloud, discarding some internal triggering due to its large size. \citet{clark09} suggest that star formation in the W51 complex is on going for the last 3 Myr at least, due to the identification of [OMN2000] LS1 as an extreme P Cygni Supergiant. On one hand, \citet{ginsb16} detected dozens of massive stars still in HC\ion{H}{2} stage, suggesting that they may be still in accretion phase with ages below 1 Myr and may represent the set of the youngest objects in the cluster. On the other hand, the fact there are sources that can be identified observing their photospheric lines is an indication that also there are sources older than 1 Myr and they may represent some of the oldest objects in W51A. The co-existence of two populations with such age spread in the same cloud is an indication of an inhomogeneous star formation history.

Our data do not suggest hierarchical formation or some age/mass gradient that could favor triggering from any source. Our conclusions also support the scenario of still ongoing multi-seeded star formation.

\section{Summary}\label{summary}

We obtained medium resolution $K$-band spectra ($R$\,=\,5,200) of a sample of eight NIR counterparts of UC\ion{H}{2} regions (out of nine candidates) in W51A, the most active region in the massive star formation complex W51. The sample includes the NIR counterpart of W51e, the brightest source at both infrared and radio wavelengths, according to the list presented by \citet{meh94}.

We detected photospheric lines in six sources, leading us to classify them as massive stars in the range O8-B3. When compared to the classification derived from their radio flux, we found similar spectral types, except for sources W51a and W51c$_1$, which we assigned a somewhat later spectral type. We could only detect nebular emission lines of H and He often observed in embedded MYSOs for W51e and W51e$_2$. Typical photospheric lines of late type stars were detected in the last object, leading us to classify it as a K7III type star, which excludes the source as the ionizing source of W51g.

We derived an average spectrophotometric distance to W51A including four O-type stars identified by \citet{lys08} and another three sources from \citet{bik19} to the five massive stars identified in this work. We adopted the reddening law of \citet{dami16} to estimate the extinction towards the line-of-sight of each source. The distance obtained in this study is $d$\,=\,4.80\,$\pm$\,1.27\,kpc, in good agreement with radio kinematic distances and compatible with distances obtained by trigonometric parallaxes of masers, within the uncertainties involved. The new extinction law resolves the discrepancy of a shorter distance as derived by \cite{lys08}.

We further analyzed the HR diagram showing the sample of 12 massive objects associated with the W51A. The position of the sources suggests that W51A is somewhat older than previous estimates \citep[e.g.][$<$1 Myr]{oku00}.

\acknowledgments

The authors thank the anonymous referee for the comments and suggestions that improved the clarity of this paper. Also, we would like to acknowledge the Hawaiian people for letting astronomers unveil the secrets of the universe from a sacred place like Maunakea. We are most fortunate to have the opportunity to conduct observations from this mountain, mahalo! All of the authors are deeply indebted to our colleague and friend, Peter Conti for years of support, mentoring, and collaboration. CLB wishes to thank FAPESP for funding the research on massive star formation through more than a decade (e.g. 2006/02467-0 and 2018/06695-4) and NOAO (now NOIRLab) for the warm reception while visiting its facilities to discuss the present research. The work of FN is supported by NOIRLab, which is managed by the Association of Universities for Research in Astronomy (AURA) under a cooperative agreement with the National Science Foundation. FN acknowledges FAPESP for financial support under grant 2017/18191-8. AD thanks FAPESP for support through the grant 2011/51680-6.

This research has made use of the VizieR catalogue access tool, CDS, Strasbourg, France (DOI : 10.26093/cds/vizier). The original description of the VizieR service was published in 2000, A\&AS 143, 23

\vspace{5mm}
\facilities{Gemini:Gillette (NIFS, ALTAIR)}

\bibliographystyle{aasjournal}
\bibliography{references}

\end{document}